\documentclass[sigconf,edbt]{acmart-edbt2021}

\def\BibTeX{{\rm B\kern-.05em{\sc i\kern-.025em b}\kern-.08em
    T\kern-.1667em\lower.7ex\hbox{E}\kern-.125emX}}

\usepackage{booktabs} 

\usepackage{tabularx}
\usepackage{multirow}
\usepackage{url}

\setcopyright{rightsretained}

\acmDOI{}

\acmISBN{978-3-89318-084-4}

\acmConference[EDBT 2021]{24th International Conference on Extending Database Technology (EDBT)}{March 23-26, 2021}{Nicosia, Cyprus} 
\acmYear{2021}

\begin{document}
\title{Coining goldMEDAL: A New Contribution to\\Data Lake Generic Metadata Modeling}


\author{\'Etienne Scholly}
\affiliation{%
  \institution{Universit\'e de Lyon, Lyon 2, \\UR ERIC \& BIAL-X}
  \streetaddress{5 avenue Pierre Mendès France}
  \city{Lyon} 
  \country{France} 
  \postcode{69676}
}
\email{etienne.scholly@bial-x.com}

\author{Pegdwendé N. Sawadogo}
\affiliation{%
  \institution{Universit\'e de Lyon, Lyon 2, \\UR ERIC}
  \streetaddress{5 avenue Pierre Mendès France}
  \city{Lyon}
  \country{France} 
  \postcode{69676}
}
\email{pegdwende.sawadogo@univ-lyon2.fr}

\author{Pengfei Liu}
\affiliation{%
  \institution{Universit\'e de Lyon, Lyon 2, \\UR ERIC}
 \city{Lyon} 
  \state{France} 
}
\email{pengfei.liu@eric.univ-lyon2.fr}

\author{Javier A. Espinosa-Oviedo}
\affiliation{%
  \institution{Universit\'e de Lyon, Lyon 2, \\UR ERIC-LAFMIA lab }
  \city{Lyon} 
  \state{France} 
  \postcode{69676}
}
\email{javier.espinosa@imag.fr}

\author{Cécile Favre}
\affiliation{%
  \institution{Universit\'e de Lyon, Lyon 2, \\UR ERIC}
  \city{Lyon} 
  \state{France} 
}
\email{cecile.favre@univ-lyon2.fr}

\author{Sabine Loudcher}
\affiliation{%
  \institution{Universit\'e de Lyon, Lyon 2, \\UR ERIC}
  \city{Lyon} 
  \state{France} 
}
\email{sabine.loudcher@univ-lyon2.fr}

\author{Jérôme Darmont}
\affiliation{%
  \institution{Universit\'e de Lyon, Lyon 2, \\UR ERIC}
  \city{Lyon} 
  \state{France} 
}
\email{jerome.darmont@univ-lyon2.fr}

\author{Camille Noûs}
\affiliation{%
  \institution{Universit\'e de Lyon, Lyon 2, \\Laboratoire Cogitamus}
  \city{Lyon} 
  \state{France} 
}
\email{camille.nous@cogitamus.fr}

\renewcommand{\shortauthors}{}

\begin{abstract}

    The rise of big data has revolutionized data exploitation practices and led to the emergence of new concepts. Among them, data lakes have emerged as large heterogeneous data repositories that can be analyzed by various methods. An efficient data lake requires a metadata system that addresses the many problems arising when dealing with big data. In consequence, the study of data lake metadata models is currently an active research topic and many proposals have been made in this regard. 
    However, existing metadata models are either tailored for a specific use case or insufficiently generic to manage different types of data lakes, including our previous model MEDAL. In this paper, we generalize MEDAL's concepts in a new metadata model called goldMEDAL. Moreover, we compare goldMEDAL with the most recent state-of-the-art metadata models aiming at genericity and show that we can reproduce these metadata models with goldMEDAL's concepts. 
    As a proof of concept, we also illustrate that goldMEDAL allows the design of various data lakes by presenting three different use cases.
    
    

\end{abstract}

%
%



\maketitle

\section{Introduction}

    While the big data revolution has shaken up the entire field of data management and analytics, new concepts have emerged to meet these new challenges. Data lakes belong to such new concepts. First introduced by James Dixon, a data lake is a vast repository of raw and heterogeneous data from which various analyses can be performed ~\cite{Dixon2010}. 
    Data lakes quickly gained popularity and several teams started to address research issues~\cite{Madera2016, Miloslavskaya2016}. 
    A key one is efficient metadata management for avoiding data lakes to turn into unexploitable data swamps~\cite{Inmon2016,Suriarachchi2016,Khine2017,Quix2018,Sawadogo2020}. 
    
    However, most metadata management proposals in the literature~\cite{Hai2016,Beheshti2018,Mehmood2019}, and their associated implementations, give few details on the way data are conceptually organized and are thence hardly reusable. 
    Thus, other researchers proposed more theoretical approaches named metadata models. Such approaches aim to provide detailed guidelines to metadata system design, while being generic, i.e., flexible and adaptable to many use cases.
    Yet, data lake generic metadata modeling is still an open research issue. A feature-based assessment indeed shows that none of the existing metadata models is generic enough, including our own MEtadata model for DAta Lakes (MEDAL)~\cite{Sawadogo2019Medal}.
    
    To address this genericity issue, we introduce goldMEDAL, a revision of our MEDAL model. 
    We define goldMEDAL through a classical three-level modeling process (i.e., conceptual, logical and physical). We choose a formal representation to avoid ambiguity but also provide a UML representation for readability. The logical level is a translation of the concepts using graph theory. Eventually, we describe three different physical models as proofs of concept. 
    Furthermore, to highlight goldMEDAL's genericity, we show that the concepts of our metadata model help model state-of-the-art metadata models from the literature. 

    The remainder of this paper is organised as follows. Section~\ref{sec:models} reviews and discusses existing data lake metadata models. Section~\ref{sec:goldmedal} presents goldMEDAL's conceptual and logical models. Section~\ref{sec:assessment} illustrates how goldMEDAL generalises other data lake metadata models and how it can be used to implement different data lakes. Finally, Section~\ref{sec:concl} concludes this paper and hints at future research.  

\section{Related Works}
\label{sec:models}
    
    Metadata management plays a vital role in data lakes. Indeed, in the absence of a fixed schema, data querying and analyses depend on an efficient metadata system. Several approaches help manage metadata in data lakes. However, only a few of them provide enough detail to ensure reusability. We refer to them as metadata models. In this section, we 
    review state-of-the-art
    metadata models (Section~\ref{sec:presentation.models}) and compare them with respect to genericity  
    (Section~\ref{sec:comparison.models}).
    
    \subsection{Metadata Models for Data Lakes}
    \label{sec:presentation.models}
    
    GEMMS (Generic and Extensible Metadata Management System) is a pioneer generic metadata model for data lakes \cite{Quix2016A}. GEMMS features two abstract entities: {\em data file} and {\em data unit}. A data file represents a generic data source. A data unit represents an identifiable data element inside a data source. Each data file is composed of a set of data units (e.g., a spreadsheet file is composed of a set of sheets). Data files and data units can be enriched with atomic or complex metadata values. However, GEMMS requires information on data structure to operate. Thus, making it unsuitable for working with unstructured data. 
    
    {\em Ground} is another generic metadata model \cite{Hellerstein2017} 
    that can be used for modeling metadata in data lakes (although not specifically designed for that). Ground tracks {\em data context} (metadata) at three levels: 1) metadata properties, 2) data usage history and 3) data versioning. Although more extensive than GEMMS, Ground (as well as GEMMS) does not take in charge data linkage even though this type of metadata has been identified as relevant in data lakes~\cite{Farrugia2016,Sawadogo2019Medal}.
    
    Based on GEMMS' data file and data units concepts, 
    The model of Diamantini et al. adds {\em similarity links} between data units to indirectly link data files \cite{Diamantini2018}. 
    However, their model  does not include important metadata such as data versioning and usage tracking as compared to Ground.
    
    Similar to Diamantini et al., Ravat and Zhao propose a model where each data file can be associated with atomic and complex metadata \cite{ravat2019metadata}, including metadata properties, data history and links with other data files. The main contribution of this model is the notion of {\em zone} metadata. Many data lake architectures consider the existence of zones (e.g., raw data zone, processed data zone) \cite{Giebler2019,ravat2019metadata}. Zone metadata specifies the zones where data is located. 
    However, Ravat and Zhao's model cannot simultaneously represent different data granularity levels as previous models do~\cite{Quix2016A,Diamantini2018}. 
    
    MEDAL represents data through three main concepts: {\em data objects}, {\em representations} and {\em versions}~\cite{Sawadogo2019Medal}. Data objects correspond to GEMMS' data files. Representations correspond to the result of transformed objects. Versions represent objects updates. Both, representations and versions, are materialized in the data lake. Thus, MEDAL gives alternative ways to track data linkage and zone metadata through the concepts of versions and representations, respectively. MEDAL also supports linkage metadata through categorizations and similarity links. 
    However, MEDAL does not support multiple data granularity levels either.
    
    Finally, HANDLE (Handling metAdata maNagement in Data LakEs), uses the generic concept of {\em data entity} to represent both, data files and parts of data files, which helps HANDLE support any granularity level~\cite{Eichler2020}.
    In HANDLE, each data entity is associated with tags that represent zones, granularity levels or categorizations. HANDLE can also connect data entities together through containment links (e.g., between a table and a tuple). HANDLE provides concepts that subsume most of the concepts of the previous metadata models. 

    \subsection{Genericity of Metadata Models}
    \label{sec:comparison.models}
    
    A generic metadata model should adapt to any data lake use case. As each use case requires specific metadata management features, we consider that the most abundant features a metadata model supports, the most generic it is.  Therefore, features are a suitable way to compare metadata models.
    
    
    To the best of our knowledge, there exist two feature-based comparisons of data lake metadata models in the literature. We introduced six relevant features: semantic enrichment, data indexing, data polymorphism, data versioning, link generation and usage tracking \cite{Sawadogo2019Medal}; while Eichler et al. identified three other features: metadata properties, zone metadata and the support of multiple granularity levels \cite{Eichler2020}.
    
    Considering that both the above sets of features are relevant, we propose to combine them for comparing the genericity of metadata models. Beyond simply unioning features, we merge data polymorphism with zone metadata, as these features both refer to the same concept. 
    We also split link generation in two new features, namely similarity links and categorization, because some metadata models support only one of them. 
    Eventually, we omit data indexing in this comparison, considering that indexing does not actually induce metadata modeling issues. Although indexing is definitely relevant to assess metadata systems  \cite{Sawadogo2019Medal}, this feature seems less suited to metadata models. 
    
    All in all, we obtain a list of eight features that can serve to compare data lake metadata models and evaluate their genericity.
    \begin{enumerate}
        \item Semantic enrichment
        \item Data polymorphism/multiple zones
        \item Data versioning
        \item Usage tracking
        \item Categorization
        \item Similarity links
        \item Metadata properties
        \item Multiple granularity levels
    \end{enumerate}
    
    Table~\ref{tab:comparison} highlights the features supported by all the models reviewed in Section~\ref{sec:presentation.models}. It shows that none of them support all the features we identify. 

    \begin{table*}[ht]
        \centering
        \caption{Features supported by data lake metadata models}
        \label{tab:comparison}
        \begin{tabular}{r c c c c c c c}
        \toprule 
            \textbf{Features~$\downarrow$ $\backslash$ Models~$\rightarrow$} & GEMMS & Ground & Diamantini et al. & Ravat \& Zhao & MEDAL &  HANDLE & goldMEDAL \\  \midrule 
            
            
            Semantic enrichment  
            & \checkmark
            & \checkmark 
            & \checkmark 
            & \checkmark 
            & \checkmark 
            & \checkmark
            & \checkmark
            \\ \midrule
             
            
            Polymorphism/multiple zones  
            &  
            &  
            & \checkmark 
            & \checkmark 
            & \checkmark 
            & \checkmark
            & \checkmark
            \\ \midrule
        
            Data versioning  
            &  
            & \checkmark 
            &  
            & \checkmark 
            & \checkmark 
            &
            & \checkmark
            \\ \midrule
            
            Usage tracking  
            &  
            & \checkmark 
            &  
            & \checkmark 
            & \checkmark 
            & \checkmark
            & \checkmark
            \\ \midrule
            
            Categorization  
            & \checkmark 
            & \checkmark 
            &  
            & \checkmark 
            & \checkmark 
            & \checkmark
            & \checkmark
            \\ \midrule 
            
            Similarity links 
            &  
            &  
            & \checkmark 
            & \checkmark 
            & \checkmark 
            & \checkmark
            & \checkmark
            \\ \midrule
        
            Metadata properties 
            & \checkmark 
            & \checkmark 
            &  
            & \checkmark 
            & \checkmark 
            & \checkmark
            & \checkmark
            \\ \midrule
            
            Multiple granularity levels 
            & \checkmark 
            &  
            & \checkmark 
            &  
            &  
            & \checkmark
            & \checkmark
            \\ \midrule
            
            \textbf{Total}
            & 4/8 
            & 5/8
            & 4/8
            & 7/8
            & 7/8
            & 7/8
            & 8/8
            \\ \bottomrule
            
        \end{tabular}
    \end{table*}
    


\section{goldMEDAL Metadata Model}
\label{sec:goldmedal}

        
    Section~\ref{sec:presentation.models} establishes that, of the eight criteria used to compare data lake metadata models, none ticked all the boxes. In this section, we thoroughly describe goldMEDAL, a substantial evolution of MEDAL that generalizes its concepts while addressing all the features identified in Section~\ref{sec:comparison.models}. 
      
    A metadata model can be expressed ``in the form of an explicit schema, a formal definition, or a textual description'' \cite{Eichler2020}. In this paper, we choose a formal approach for the sake of precision. Yet, for the sake of readability and communication with possibly non-computer scientists, we also provide a semi-formal UML model.
    Moreover, we use a conventional data modeling approach that leverages a conceptual, a logical and a physical model, to demonstrate the actual implementation process of our metadata model.
        
    Section~\ref{sec:goldmedal.conceptual} presents goldMEDAL's formal and semi-formal conceptual models. 
    Section~\ref{sec:goldmedal.logical} details the translation of goldMEDAL's concepts into a logical, graph-based model. For the sake of clarity, the examples we use are the same examples in both sections, i.e., examples at the conceptual level are translated at the logical level. Eventually, example physical models, i.e., metadata models actually implemented in data lakes with goldMEDAL, are presented in Section~\ref{sec:assessment.physical}.
    
    \subsection{Conceptual Model}
    \label{sec:goldmedal.conceptual}

    
    In MEDAL, data items were considered either as \textit{raw data}, or as \textit{versions} or \textit{representations} derived from raw data. The concepts of version and representation were used to express updated and transformed data, respectively. 
    While modeling metadata for various data lakes, we found that more data items were possible, e.g., temporal representations. Thus, we decided to generalize any such concepts into a global concept named \textbf{data entity} in goldMEDAL. 
    
    Accordingly, we also generalized in goldMEDAL:
    \begin{itemize}
        \item \textit{update} and  \textit{transformation} operations that  served to track the lineage of representations and versions, respectively, as well as \textit{parenthood relationships} that express fusion operations, into the concept of \textbf{process};
        \item \textit{similarity links} into the global concept of \textbf{link}.
    \end{itemize}

    Eventually, we retained in goldMEDAL the MEDAL concept of \textbf{grouping}, which notably allows multiple data granularity levels.
    
    
    All the main goldMEDAL concepts (data entity, grouping, link and process) are characterized by attributes or properties that constitute their internal metadata. 
    
        
    \subsubsection{Data Entity}
    
        Data entities are the basic units of our metadata model. They are flexible in terms of data granularity. For example, a data entity can represent a spreadsheet file, a textual or semi-structured document, an image, a database table, a tuple or an entire database. 
        The introduction of any new element in the data lake leads to the creation of a new data entity.
        
        \begin{definition}
        The set of data entities is denoted $\mathcal{E} = \{ e_i \}_{i \in \mathbb{N}^*}$. 
        \end{definition}

    \subsubsection{Grouping}
    \label{sec:grouping}
    
        A grouping is a set of groups; a \textbf{group} brings together data entities based on common properties. For example, the raw and preprocessed data zones common in data lake architectures are the groups of a zone grouping. Another example is a grouping of textual documents according to the language of writing.
        
        \begin{definition}
        The set of groupings is denoted $\mathcal{G} = \{ G_j \}_{j \in \mathbb{N}^*}$, with $G_j = \{\Gamma_{jk}\}_{k \in \mathbb{N}^*}$ and $\Gamma_{jk}  \subseteq \mathcal{E}$ is a group.  
        \end{definition}
        
        \begin{example} \label{ex:grouping}
        To get back to our previous examples, $\mathcal{G} = \{ G_1, G_2 \}$. $G_1 = \{\Gamma_{11}, \Gamma_{12}\}$ is the zone grouping, with $\Gamma_{11}$ and $\Gamma_{12}$ being the raw data and processed data zones, respectively. $G_2 = \{\Gamma_{21}, \Gamma_{22}\}$ is the language grouping, with $\Gamma_{21}$ and $\Gamma_{22}$ the groups corresponding to French and English languages, respectively. 
        Note that the groupings $G_j$ are deliberately not partitions of $\mathcal{E}$. Thus, a bilingual French-English document can belong to both groups $\Gamma_{21}$ and $\Gamma_{22}$.
        \end{example}

    \subsubsection{Link}
    \label{sec:link}
    
        Links are used to associate either data entities with each other or groups of data entities with each other. They can be oriented or not. They allow the expression of, e.g., simple similarity links between data entities or hierarchies between groups. For example, a temporal hierarchy month $\rightarrow$ quarter would have the months of January, February and March linked to the first quarter of a given year.

        \begin{definition}
        The set of links is denoted $\mathcal{L} = \{ l_m \}_{m \in \mathbb{N}^*}$, with either:
        \begin{itemize}
            \item $l_m : \mathcal{E} \rightarrow \mathcal{E}$,
            \item $l_m : G_j \rightarrow G_{j'}$ and $j \neq j'$.
        \end{itemize}
        \end{definition}
        
        \begin{example} \label{ex:link}
        Let us elaborate the sample hierarchy month $\rightarrow$ quarter. Let   
        $G_3 = \{ Jan, Feb, ..., Dec \}$ a grouping of data entities per month and $G_4 = \{ Q1, Q2, Q3, Q4 \}$ be a grouping of quarters in a year. Now, let us make explicit some data entities and their groups: $Jan = \{ e_1, e_2 \}$, $Fev = \{ e_3 \}$, $Mar = \{ e_4 \}$; $Q1 = \{ e_1, e_2, e_3, e_4 \}$. Link $l_1$ materializes the hierarchical link between groups $G_3$ and $G_4$:  $Jan \xrightarrow[l_1]{} Q1, Feb \xrightarrow[l_1]{} Q1, Mar \xrightarrow[l_1]{} Q1$. Inversely, $Q1 \xrightarrow[l^{-1}_1]{} \{ Jan, Feb, Mar \}$. 
        
        A functional notation may also be used: $l_1 (Jan) = Q1$, $l_1 (Feb) = Q1$, $l_1 (Mar) = Q1, l_1^{-1} (Q1) = \{ Jan, Feb, Mar \}$. Also note that $Q1 = Jan \cup Feb \cup Mar$.
        \end{example}

    \subsubsection{Process}
    \label{sec:process}

        A process refers to any transformation applied to a set of data entities that produces a new set of data entities.
        
        \begin{definition}
        The set of processes is denoted $\mathcal{P} = \{ P_n \}_{n \in \mathbb{N}^*}$, with $P_n = \{I_n, O_n \}$, $I_n \subseteq \mathcal{E}$ the set of input data entities of $P_n$ and $O_n$ the set of output data entities that is integrated into $\mathcal{E}$ ($\mathcal{E} \leftarrow \mathcal{E} \cup O_n$).
        \end{definition}
        
        
        \begin{example} \label{ex:process}
        Process $P_1$ 
        splits a set of textual documents $D \subseteq \mathcal{E}$ into a 
        set of text fragments $F \subseteq \mathcal{E}$. Here, $I_1 = D$ and $O_1 = F$. 
        \end{example}

    \subsubsection{UML model}
        
\begin{figure}[h]
            \centering
            \includegraphics[width=\linewidth]{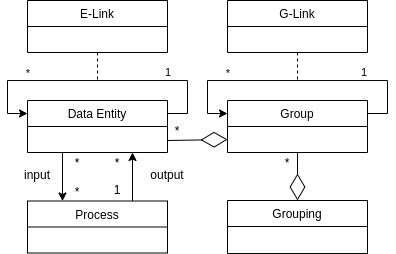}
            \caption{UML class diagram of goldMEDAL}
            \label{fig:uml}
        \end{figure}

    Figure~\ref{fig:uml} features goldMEDAL's conceptual model as a UML class diagram. All the concepts of goldMEDAL, including group, are modeled as classes (data entity, grouping, group and process) or association classes (entity link and group link, which are labeled E-Link and G-Link in Figure~\ref{fig:uml}, respectively).
    
    Eventually, although they are not depicted in Figure~\ref{fig:uml}, all classes and association classes bear attributes that model metadata properties. These attributes may be of any type, including lists, and of course vary with respect to use cases.
        
    \subsection{Logical Model}
    \label{sec:goldmedal.logical}
    
        
        As MEDAL and HANDLE did, though at the physical level, we choose to design goldMEDAL's logical model 
        as a graph, which 
        is particularly well-suited to depict relationships between different concepts. 
        
        Thus, in this section, we translate the concepts defined in Section~\ref{sec:goldmedal.conceptual} into graph nodes, edges and hyperedges, using the same indices, e.g., $i, j, k...$ Moreover, we illustrate the translation with 
        the examples used at the conceptual level. 
        Finally, we also propose a graphic illustration of goldMEDAL's logical model. 
    
        \subsubsection{Translation of Data Entity}
        
            Data entities are modeled by nodes that carry attributes.
            
            \begin{definition}
            The set of nodes is denoted $\mathcal{N} = \{n_i\}_{i \in \mathbb{N}^*}$. Each node $n_i \in \mathcal{N}$ carries attributes.
            \end{definition}
            
            \begin{example}
            A PDF file stored in the data lake can be represented by a node $n_1$. 
            \end{example}
            
        \subsubsection{Translation of Grouping}
        
        
            A group is represented by a non-oriented hyperedge, i.e., an edge that can link more than two nodes. A grouping is modeled by a set of hyperedges.
            
            \begin{definition}
            A hyperedge (a group) is denoted $\theta_{jk} \subseteq \mathcal{N}$, with $j, k \in \mathbb{N}^*$. Any $\theta_{jk}$ carries attributes. 
            \end{definition}
            \begin{definition}
            The set of hyperedges of grouping $j$ is denoted $H_j = \{ \theta_{jk} \}$ and carries attributes. The set of hyperedge sets (set of groupings) is denoted $\mathcal{H}$. 
            \end{definition}

            \begin{example}
            \label{ex:grouping_graph}
            Let us translate Example~\ref{ex:grouping}. $\mathcal{H} = \{ H_1, H_2 \}$. $H_1 = \{\theta_{11}, \theta_{12}\}$ is the set of hyperedges representing the zone grouping, with $\theta_{11}$ and $\theta_{12}$ the hyperedges representing the raw data and processed data zones, respectively. $H_2 = \{\theta_{21}, \theta_{22}\}$ is the set of hyperedges representing the language grouping, with $\theta_{21}$ and $\theta_{22}$ the hyperedges representing the groups corresponding to French and English languages, respectively. 
            \end{example}

        \subsubsection{Translation of Link}
            
            Links may model relationships between either data entities (nodes) or groups (hyperedges). They are modeled by edges.
            
            \begin{definition}
            The set of edges is denoted $\mathcal{A} = \{a_m\}_{m \in \mathbb{N}^*}$, with any $a_m$ being either:
            \begin{itemize}
                \item an edge, oriented or not, connecting two nodes. Then, $a_m = (n_i, n_{i'})\in \mathcal{N}^2$; 
                \item an oriented edge connecting two hyperedges. Then, $a_m = (\theta_{jk}, \theta_{j'k'}) \in H_j \times H_{j'}$. 
            \end{itemize}
            In both cases, the edge carries attributes.
             \end{definition}
             
            \begin{example}
            \label{ex:link_graph}
            To get back to the sample hierarchy month $\rightarrow$ quarters from Example~\ref{ex:link},  
            $H_3 = \{ \theta_{Jan}, \theta_{Feb}, ..., \theta_{Dec} \}$ is a set of hyperedges representing a grouping of data entities per month.  $H_4 = \{ \theta_{Q1}, \theta_{Q2}, \theta_{Q3}, \theta_{Q4} \}$ is a set of hyperedges representing the grouping of quarters in a year. Let us make this explicit with instances. 
            $\theta_{Jan} = \{ n_1, n_2 \}$, $\theta_{Fev} = \{ n_3 \}$, $\theta_{Mar} = \{ n_4 \}$; $\theta_{T1} = \{ n_1, n_2, n_3, n_4 \}$. Edge $a_1$ materializes the hierarchical link between $H_3$ and $H_4$:  $\theta_{Jan} \xrightarrow[a_1]{} \theta_{Q1}, \theta_{Feb} \xrightarrow[a_1]{} \theta_{Q1}, \theta_{Mar} \xrightarrow[a_1]{} \theta_{Q1}$. Inversely, $\theta_{Q1} \xrightarrow[a^{-1}_1]{} \{ \theta_{Jan}, \theta_{Feb}, \theta_{Mar} \}$.
            \end{example}
            
            
        \subsubsection{Translation of Process}
        
            A process is modeled by an oriented hyperedge.
            
            \begin{definition}
            The set of oriented hyperedges modeling processes is denoted $\mathcal{Q} = \{ \Pi_n \}_{n \in \mathbb{N}^*}$, with  
            $\Pi_n = \{ \Upsilon_n, \Omega_n \}$, $\Upsilon_n \subseteq \mathcal{N}$ being the set of input nodes of $\Pi_n$ 
            and $\Omega_n$ the a set of output nodes integrated to $\mathcal{N}$ ($\mathcal{N} \leftarrow \mathcal{N} \cup \Omega_n$). Any $\Pi_n$ carries attributes. 
            \end{definition}
            
            \begin{example}
            \label{ex:process_graph}
            $\Pi_1 = \{ \Upsilon_1, \Omega_1 \}$ is an oriented hyperedge representing the process of splitting a set of textual documents (Example~\ref{ex:process}) represented by the set of nodes $N_D \subseteq \mathcal{N}$, into a set of text fragments represented by the set of nodes $N_F \subseteq \mathcal{N}$. Then, $\Upsilon_1 = N_D$ and $\Omega_1 = N_F$. 
            \end{example}

            
        \subsubsection{Sample Graph Representation}
        
\begin{figure*}[h]
           \centering
           \includegraphics[width=\textwidth]{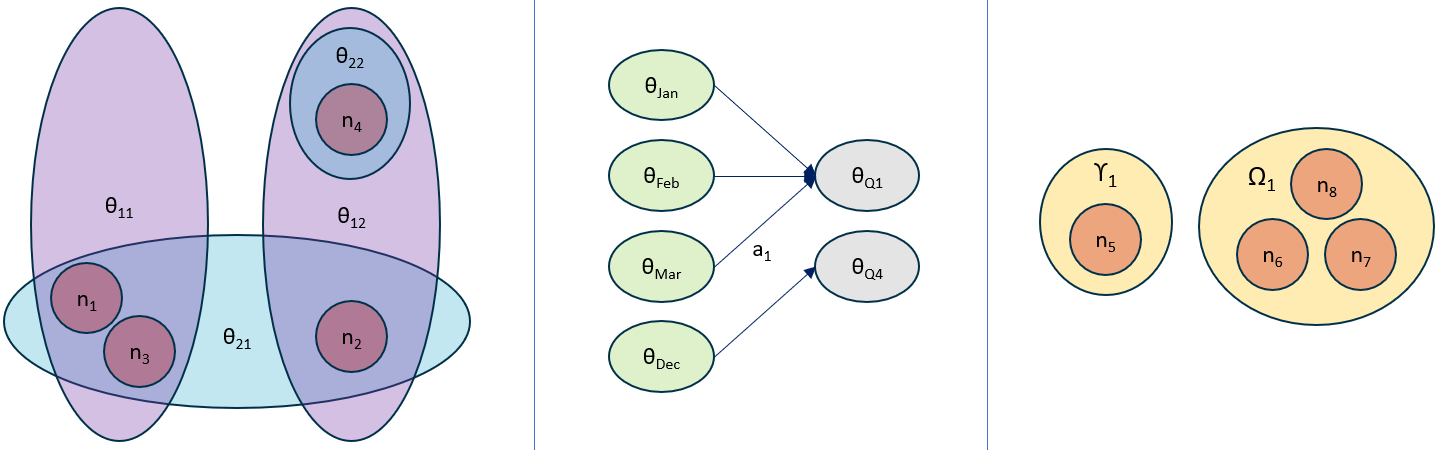}
            \caption{Sample goldMEDAL graph logical model}
            \label{fig:graph}
        \end{figure*}

        Figure~\ref{fig:graph} provides a sche-matic representation of the examples above. Let us introduce eight data entity nodes $\{ n_i \}_{i \in [1,8]}$  colored in orange.
        
        Example~\ref{ex:grouping_graph} is depicted on the left-hand side of Figure~\ref{fig:graph}. Groups of $H_1$ are colored in purple, while $H_2$'s are blue. We can see that $n_1$ and $n_3$ belong to the raw data group $\theta_{11}$, while $n_2$ and $n_4$ are in the processed data group $\theta_{12}$. Moreover, $n_1$, $n_2$ and $n_3$ are in the French language group $\theta_{21}$, and $n_4$ is in the English language group $\theta_{22}$.
        
        Example~\ref{ex:link_graph} is represented at the center of Figure~\ref{fig:graph}. Groups of $H_3$, namely $\theta_{Jan}$, ..., $\theta_{Dec}$ are colored in green and groups of $H_4$ ($\theta_{Q1}$, ..., $\theta_{Q4}$) are colored in grey. Hyperedge $a_1$ connects groups of $H_3$ to $H_4$'s.
        
        Finally, Example~\ref{ex:process_graph} is depicted on the right-hand side of Figure~\ref{fig:graph}. $n_5$ is a textual document split in fragments $n_6$, $n_7$ and $n_8$. $\Pi_1$'s input and output $\Upsilon_1$ and $\Omega_1$, respectively, are colored in yellow. 
        
\section{goldMEDAL Assessment}
\label{sec:assessment}
    
    In this section, we discuss goldMEDAL's genericity. To this end, we show in Section~\ref{sec:assessment.modeling} that all three most complete metadata models can be modeled with goldMEDAL. In Section~\ref{sec:assessment.physical}, we present our ongoing implementation work of goldMEDAL on distinct use cases. 

\subsection{Comparison of State-of-the-Art Metadata Models with goldMEDAL}
\label{sec:assessment.modeling}

    To evaluate goldMEDAL's genericity, we compare it with the three metadata models 
    that are both the most recent and the most complete among metadata models, i.e., MEDAL, Ravat and Zhao's and HANDLE  (Section~\ref{sec:comparison.models}).
    
   For each comparison, we use a two-column table. The first column lists goldMEDAL's concepts, and the second column the corresponding concepts of the compared model. When any concept does not have an equivalent, it is marked with ``---''.

    \subsubsection{MEDAL vs. goldMEDAL}
    
        goldMEDAL's four main concepts help generalize all of MEDAL's concepts (Table~\ref{tab:goldmedal-medal}). Data entity generalizes the concepts of version and representation. Grouping generalizes the concepts of object and grouping (in the sense of MEDAL). Link generalizes the concepts of similarity link. Finally, process generalizes transformation, update and parenthood relationship.
        
        \begin{table}[!ht]
            \centering
            \caption{goldMEDAL and MEDAL concepts}
            \label{tab:goldmedal-medal}
            \begin{tabular}{ll}
                \toprule
                \textbf{\begin{tabular}{@{}l@{}}goldMEDAL \end{tabular}} & \textbf{MEDAL}  \\ \midrule
                Data entity & Version, Representation 
                \\ \midrule
                Grouping & Object, Grouping 
                \\ \midrule
                Link & Similarity link
                \\ \midrule
                Process & 
                Update, Transformation,\\ & Parenthood relationship
                \\ \bottomrule
            \end{tabular}
        \end{table}
    
        Note that we do not mention in this comparison global metadata existing in MEDAL. We indeed consider that elements such as logs or indexes mostly induce implementation rather than metadata modeling issues.

        Yet, other forms of global metadata, namely semantic resources such as thesauruses and ontologies, can definitely be modeled with goldMEDAL using the node, grouping and link concepts. 
    
    \subsubsection{Ravat and Zhao's Metadata Model vs. goldMEDAL}
    
        gold-MEDAL can handle nearly all concepts of Ravat and Zhao's metadata model~\cite{ravat2019metadata} (Table~\ref{tab:goldmedal-zhao}). Data entity generalizes the concept of dataset and all its subclasses, such as Datalake\_Datasets or Source \_Datasets. Grouping generalizes the concepts of keyword. Finally, link and process directly correspond to relationship and process, respectively.
        
        \begin{table}[!ht]
            \centering
            \caption{goldMEDAL and Ravat \& Zhao concepts}
            \label{tab:goldmedal-zhao}
            \begin{tabular}{ll}
                \toprule
                \textbf{goldMEDAL} & \textbf{Ravat \& Zhao}
                \\ \midrule
                Data entity & Dataset, Subclass
                \\ \midrule
                Grouping & Keyword
                \\ \midrule
                Link & Relationship 
                \\ \midrule
                Process & Process
                \\ \midrule
                --- & User, Access
                \\ \bottomrule
            \end{tabular}
        \end{table}
        
        However, two concepts of Ravat and Zhao's metadata model, namely user and access, have no explicit equivalent in goldMEDAL, though they could be classified as global metadata. 
        Users and accesses can indeed be modeled as data entities and processes, respectively. 

    \subsubsection{HANDLE vs. goldMEDAL}
    
        goldMEDAL can also generalize HANDLE's concepts (Table~\ref{tab:goldmedal-handle}). Data entity generalizes both data and metadata, since a data entity is a representation of data that also contains metadata properties. Grouping generalizes three concepts: Categorization, ZoneIndicator, and GranularityIndicator. Finally, process has no direct match in HANDLE, although its authors show processes can be modeled through Action metadata instances of HANDLE's categorization extension~\cite{Eichler2020}.
        
        

        \begin{table}[!ht]
            \centering
            \caption{goldMEDAL and HANDLE concepts}
            \label{tab:goldmedal-handle}
            \begin{tabular}{ll}
                \toprule
                \textbf{goldMEDAL} & \textbf{HANDLE}  \\
                \midrule
                Data entity & Data, Metadata
                \\ \midrule
                Grouping & Categorization, ZoneIndicator
                \\ 
                  &  GranularityIndicator
                \\ \midrule
                Link & Link
                \\ \midrule
                Process & ---
                \\ \bottomrule
            \end{tabular}
        \end{table}

        Handling multiple granularity levels as in HANDLE was not supported by MEDAL, so it was a design objective for goldMEDAL. Although there is no explicit granularity indicator in goldMEDAL, any data entity could have a granularity property. However, there is more efficient way by defining data entities on the finest possible granularity level. Then, coarser granularity levels are obtained with groupings. For example, if each data entity corresponds to a tuple in a relational database, then a grouping represent a set of tables.

\subsection{goldMEDAL Physical Models}
\label{sec:assessment.physical}

    
    To show that goldMEDAL can model different business issues and manage various functionalities while remaining as simple as possible, we apply our metadata model to three different use cases. 
    We also exemplify how goldMEDAL's logical model (Section~\ref{sec:goldmedal.logical}) can be translated into different physical models.

    
    \subsubsection{Public Housing Data Lake}
    
        For social landlords (agents or agencies providing social housing), the use of data is nothing new, whether through business intelligence for patrimony management or with data science methods for non-payment forecasting. 
        However, landlords are facing two main problems. On the one hand, their analyses are conducted separately: in different environments, by different individuals and with different tools. This implies that collaborative work on the same data is impossible. On the other hand, landlords know how to use their data, but have much more difficulty capturing and exploiting ``external'' data. Yet their dwellings are located in environments with their own characteristics (transportation, climate, employment rate, education, etc.), which affect the attractiveness of the dwellings. Being able to combine this external information with landlords' data would be a real asset for understanding their patrimony. 
        
        A data lake can store both ``internal'' data from social landlords as well as ``external'' data gathered on the Internet. In addition, all types of analyses can be carried out from the data lake. 
        
        \paragraph{HOUDAL (\textit{public HOUsing DAta Lake})} The data lake implemented for social landlords~\cite{scholly2019business} is based on a Web application, and thus is composed of two major parts: the front-end (or client part) is the user interface for depositing new data, for creating new metadata and for consulting existing metadata; the back-end (or server part) features various services such as an API, the metadata system, data storage, and a user management service. 

        \paragraph{HOUDAL Metadata System}
        goldMEDAL's metadata model has been implemented into the Neo4J graph database management system\footnote{\url{https://neo4j.com}}. Since Neo4J does not allow to have hyperedges, we create a node for each concept. Thus, entities, groups, groupings, links and processes translate as nodes, each bearing a label and attributes. 
        
        \paragraph{Data entities}
        The different data files that populate the data lake are data entities. They can be either raw data files sent by landlords (often in comma separated value files) or reworked data, sometimes stored in various formats such as .pkl or .RData, for Python and R analyses, respectively. Each data entity has its node labeled :ENTITY and the entity's properties, such as file name or description, are stored in the node's attributes. 

        \paragraph{Groupings for Categorizing Data Entities} With HOUDAL, users can create as many groupings as necessary, and several groups for each grouping. Data entities can be linked to zero, one or several groups for each grouping. In Neo4J, groupings are modeled by nodes carrying a :GROUPING label. Groups are also nodes, carrying both a :GROUP label and the grouping's name as a second label, in order to facilitate querying. A data entity node (resp. group node) is linked to a group node (resp. grouping node) with an edge labeled with the grouping's name (resp. :GROUPING). With groups and groupings, users can, for example, determine whether it is internal or external data, or the data refinement level (zones), and so on.
        
        \paragraph{Processes for Tracking Data Lineage} Like other goldMEDAL concepts, a process is also modeled by a node in Neo4J, bearing the :PROCESS label. A process can be a script for transforming or cleaning a data file, i.e., a data entity. If a data entity is the input of a process, there is an edge labeled :PROCESS\_IN from the entity node to the process node. Inversely, an edge labeled :PROCESS\_OUT from the process node to the entity node is created if a new data entity is generated by the process.
        
\begin{figure*}[hbt]
            \centering
            \includegraphics[width=\textwidth]{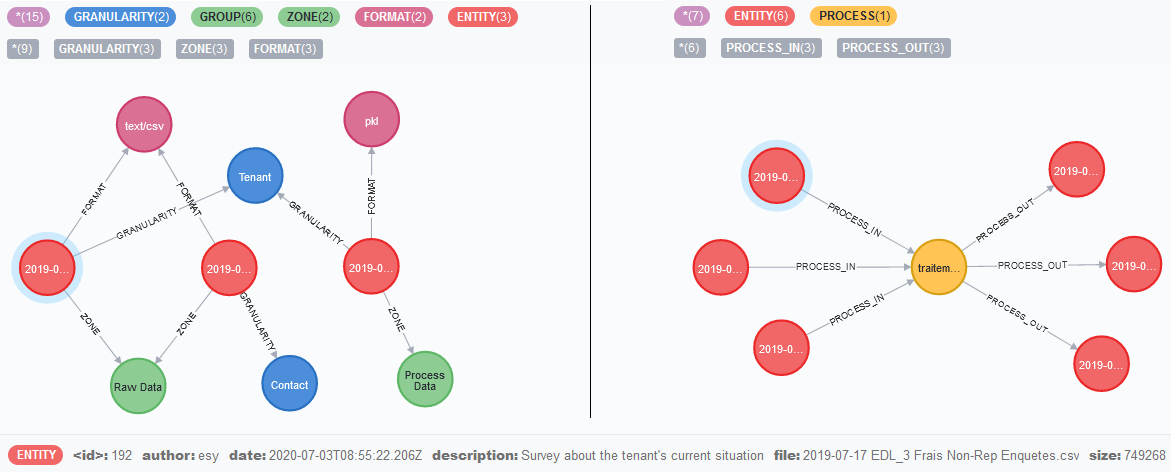}
            \caption{HOUDAL sample Neo4J metadata}
            \label{fig:houdal}
        \end{figure*}

        \paragraph{Example}
        Figure~\ref{fig:houdal} presents a sample of metadata stored in Neo4J. Data entity nodes are colored in red. On both sides of the figure, a data entity node is highlighted: some of its attributes are depicted at the bottom in grey. 
        
        The left-hand side of Figure~\ref{fig:houdal} gives an example of groupings. There are three groupings: a zone grouping, a format grouping and a granularity grouping. Each grouping has its group nodes, colored in green, purple and blue, respectively. Data entity nodes are connected to group nodes with an edge. For example, we can see that the highlighted data entity node (on the left) is a raw .csv file, and the granularity level is ``Tenant'', meaning that each line corresponds to a tenant. Note that in Neo4J, groupings are also modeled as nodes, but are not represented in this Figure.  
        
        An example of process is depicted on the right-hand side of Figure~\ref{fig:houdal}. The process node is colored in yellow. We can see that three data entity nodes are the process' input, and three data entity nodes are the process' output, meaning that they are generated by the process.
        
        HOUDAL is operational and is currently being tested by social landlords. Nevertheless, we have many areas for improvement to work on, to make the application more robust and more user-friendly. In addition, we continue to discuss with social landlords to identify new needs, which could be the subject of future work to add a new feature to our data lake. 
        
    \subsubsection{Textual and Tabular Data Lake}
    \label{sec:audal}
        

        The AUDAL data lake is motivated by researchers in management science who want to analyze the effect of servicization (i.e., the transition from supplying products to supplying services) and digitization on small and medium sized companies' economic performance \cite{ddh20}. Source data are various textual documents (annual reports, press releases, websites, social media posts) and spreadsheet files featuring qualitative (e.g., stocks) and qualitative (e.g., degree of servicization) characteristics.
        
\begin{figure*}[hbt]
        \centering
        \begin{minipage}{.5\textwidth}
        \includegraphics[width=7.7cm]{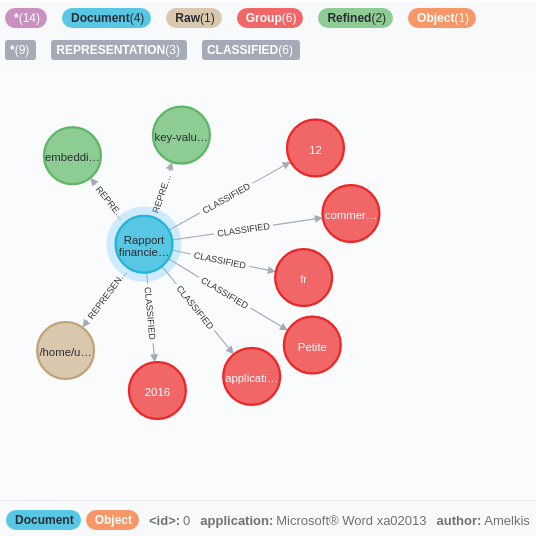} 
        \end{minipage}
        \begin{minipage}{.5\textwidth}
        \centering
        \includegraphics[width=8cm]{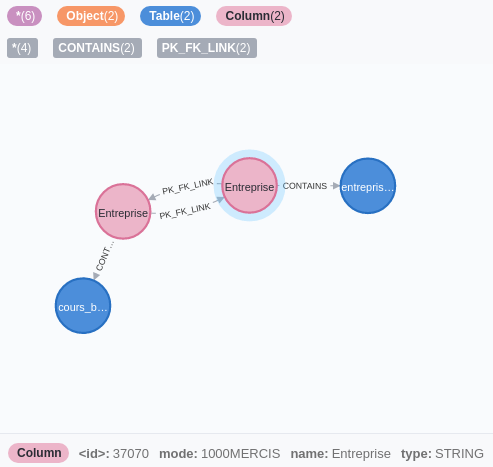} 
        \end{minipage}
        \caption{AUDAL sample Neo4J metadata}
        \label{fig:audal}
        \end{figure*}

        \paragraph{Metadata Management in AUDAL}   
        AUDAL's metadata system is architectured in three levels. The first level manages data entities. Data entities, i.e., textual documents and spreadsheet tables, are categorized as \textit{raw} and \textit{refined}. Raw tables or documents are actually pointers to the corresponding files in their original format. Raw data entities store metadata properties, in the form of Neo4J node attributes, e.g., file author(s), date of creation, etc. 
        Refined data entities are automatically generated from raw data entities. They are transformed so as to be exploited in analyses. More concretely, raw textual documents are refined into bag-of-word vectors or document embedding vectors stored in the MongoDB document-oriented database management system\footnote{\url{https://www.mongodb.com}}, and referenced from Neo4J nodes~(Figure~\ref{fig:audal}). Similarly, raw spreadsheet tables are refined in relational tables to benefit from SQL querying.

        The second level in AUDAL's metadata system handles relationships between data items. We use two kinds of relationships in accordance with goldMEDAL concepts: groupings and (similarity) links. 
        Some of the groupings relate to both tabular and textual data, e.g., groupings on the MIME type or data source. Conversely, others are relevant for only one type of data, e.g., the grouping on the language of documents. We materialize groupings in Neo4J through a set of nodes. Each grouping is a simple node with which all associated groups are linked. Then, groups are in turn linked to the corresponding data entities.
        
        We define two types of links with respect to the type of data they relate to. \textit{Document similarity links} express how much two documents are similar. These links are materialized by non-oriented edges between data entity nodes in Neo4J. 
        Similarly, we express links between tabular data with \textit{Table joinability links}. Such links (labeled \textit{PK\_FK\_LINK} in Figure~\ref{fig:audal}) actually represent some automatically detected functional dependencies between columns from different tables. In Neo4J, table joinability edges are oriented.
        
        Eventually, our model's third level is constituted of metadata used to speed up or enhance analyses. It includes indexes that allow and speed up keyword-based search on textual documents as well as spreadsheet files. These indexes are managed by ElasticSearch\footnote{\url{https://www.elastic.co}}.
        Moreover, AUDAL's metadata system also includes semantic resources, i.e., dictionaries and thesaurus. Such resources, stored in MongoDB, allow amongst other automatic query extension.
        
        
        \paragraph{Analyses with AUDAL}          
        AUDAL allows both data retrieval and content analyses. Data retrieval works in three different ways. 
        The first way exploits indexes to allow term-based queries. It is effective for both textual documents and tabular data. 
        AUDAL also provides navigation as a solution to discover data of interest. This is done by intersecting groups from different groupings. For example, such queries allow finding data from a specific source and created on a specified year. 
        Finally, data can be retrieved using relatedness, starting from a specified data object and then finding the most related data, namely similar documents or joinable tables.
                
        Content analyses are actually a way to aggregate data. In the case of textual documents, such analyses include document clustering or scoring with respect to a set of keywords and text concordance. 
        Tabular data are exploited through SQL queries, the clustering of table rows and correlation analyses between columns. 
            
    \subsubsection{Archaeological Data Lake}
    \label{sec:archaeodal}
            
        This data lake was designed during the course of the multidisciplinary project ``Hyper thesaurus and data lakes: Mine the city and its archaeological archives'' (HyperThesau) \cite{dh20,ddh20}. Let us name it ArchaeoDAL, in echo to HOUDAL and AUDAL, though it was actually never called so.
        
        Archaeological data may bear many different types, e.g., textual documents (excavation reports), images (photographs, drawings, plans...), sensor data, chemical analysis results, etc. Even structured data are often produced by various devices that are not compatible with each other. Moreover, the description of an archaeological object also differs with respect to users, usages and time. Thus, archaeologists use semantic resources such as thesauruses to interoperate data from various origins. 
                
\begin{figure*}[hbt]
            \centering
            \includegraphics[width=13cm]{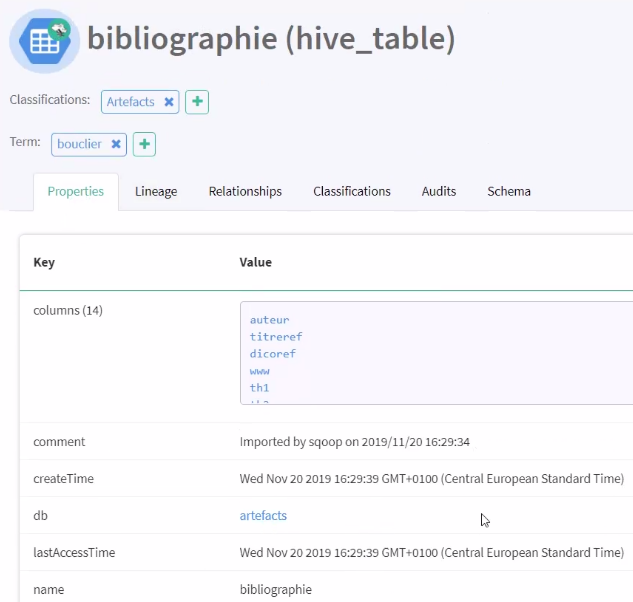}
            \caption{Sample Atlas object}
            \label{fig:atlas1}
        \end{figure*}

        \paragraph{Physical Model of Data Entities}
        The implementation of ArchaeoDAL heavily relies on the Apache ecosystem. In particular, its metadata system rests on the Atlas\footnote{\url{https://atlas.apache.org}} data governance and metadata framework. Atlas' objects match with goldMEDAL's data entities. In addition to metadata properties (in the form of key-value pairs), objects may also relate to terms from thesauruses, i.e., goldMEDAL links, and classifications, i.e., goldMEDAL groupings (Figure~\ref{fig:atlas1}). 
        
        Moreover, we exploit Atlas' object types to fulfill domain-specific requirements regarding metadata properties. For example, in the HyperThesau project, users need not only semantic metadata to understand data contents, but also geographical metadata to know where archaeological objects were discovered. The benefits of having an object type system include:
        \begin{itemize}
            \item consistency: a universal definition of metadata can avoid terminological variations that may cause data retrieval problems;
            \item flexibility: a domain-specific type system helps define specific metadata for requirements in each use case;
            \item efficiency: with a given metadata type system, it is easy to write and implement search queries. Because names and types of all metadata properties are known in advance, we can filter data with metadata predicates such as $upload\_date >$ `10/02/2016'.
        \end{itemize}
        
\begin{figure*}[hbt]
            \centering
            \includegraphics[width=12cm]{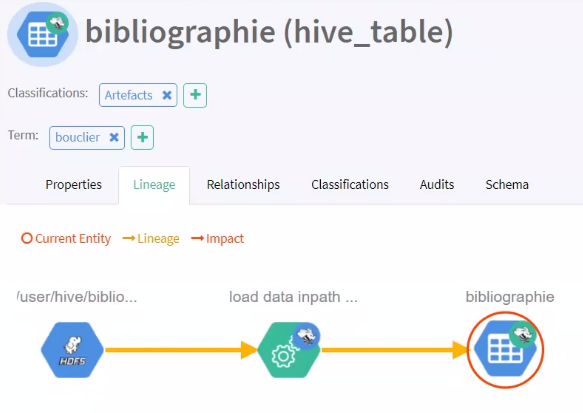}
            \caption{Sample Atlas lineage}
            \label{fig:atlas2}
        \end{figure*}

        \paragraph{Physical Model of Processes}
        Atlas also includes a nice lineage feature that helps visualize chains of processes. For instance, Figure~\ref{fig:atlas2} represents a simple ingestion process of raw data stored in HDFS into a Hive table, where objects are symbolized by blue hexagons and the process by a green hexagon.
        
        \paragraph{Thesauruses and Links}
        The HyperThesau project heavily relies on thesauruses to organize data. A thesaurus consists of a set of categories and terms that help regroup data. In Atlas' glossary, a category may have only one parent. A category without a parent is called the root category. Conversely, a category may have several subcategories or terms. A term must have a parent category but no subcategory. A term may have relationships (i.e., goldMEDAL links) with other terms, e.g., related words, synonyms, antonyms, etc. Note that it would be easy to represent ontologies or taxonomies, too. 
        
\begin{figure*}[hbt]
            \centering
            \includegraphics[width=15cm]{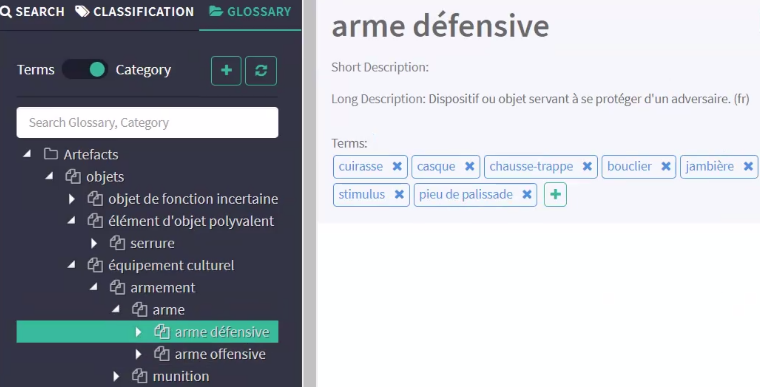}
            \caption{Sample Atlas thesaurus}
            \label{fig:atlas3}
        \end{figure*}

        Eventually, we add specific links between data nodes associated with term nodes from the thesaurus. The left-hand side of Figure~\ref{fig:atlas3} displays an excerpt of the thesaurus. Figure~\ref{fig:atlas3} also shows how a term (\textit{arme défensive}, i.e., defensive weapon) points to the corresponding metadata (short and long descriptions) and related terms.
        
\section{Conclusion}
\label{sec:concl}

    In this paper, we introduced goldMEDAL, a generic data lake metadata model. goldMEDAL is based on four main concepts: data entity, grouping, link and process, which are defined at the conceptual and logical levels. 
    These concepts interact altogether to support data lake metadata management requirements and they generalize almost all the concepts proposed in state-of-the-art metadata models : the concept of grouping supports the organization of data lakes in zones~\cite{ravat2019metadata}; groupings allow managing multiple data granularity levels 
    as in HANDLE~\cite{Eichler2020}. 
        
    Moreover, goldMEDAL supports all the features identified to compare data lake metadata models (Section~\ref{sec:comparison.models}), making it the most generic metadata model to the best of our knowledge.
       
    Another particularity of goldMEDAL is the explicit possibility of data lineage tracing with the concept of process. goldMEDAL thus manages the dynamics of data, while the most recent metadata model from the literature, HANDLE~\cite{Eichler2020}, does not natively support it.
        
    Eventually, we show as a proof of concept how goldMEDAL can be translated from conceptual and logical models to actual physical models with three different implementations of metadata models from distinct data lakes 
    that feature both structured and unstructured data. 
        
    Future research and open issues include the ``industrialization'' of data lakes, i.e., providing a software layer, connected to the metadata system, which allows non-data or non-computer scientists to transform and analyze their own data in autonomy, just as dynamic reports are prepared on top of data warehouses for the use of business (i.e, non technical) users. However, such a software layer must not become yet another black box. In consequence, we must take great care of accompanying users in their appropriation of our analysis tools, not only by training, but also by interweaving research methodologies from computer science with business practices \textit{by design}, in close collaboration with the partners.
    
    Moreover, exploiting a data lake and its metadata system may contribute to open data and open science. A well-designed data lake should indeed readily enforce the four FAIR principles\footnote{\url{https://www.go-fair.org/fair-principles/}}, i.e., findability, accessibility, interoperability and reusability. By adding an industrialization layer that allows non-data or non-computer scientist exploit the data lake, we can further improve accessibility \textit{in a non-technical way}, i.e., not only through suitable communication protocols. FAIR principles are very appealing to researchers in humanities and social sciences, as illustrated by AUDAL (management sciences; Section~\ref{sec:audal}) and ArchaeoDAL (archaeology; Section~\ref{sec:archaeodal}).
    
    Finally, to the best of our knowledge, the maintenance of data lake metadata is a completely open issue. For instance, how to manage a new categorization of metadata? How to change or transform the metadata system when it hits some limits, whether technical or functional? What if metadata become big in the sense of voluminous big data? Should obsolete data be deleted, which is contrary to the principle of data lakes, and how to ensure that the metadata accessibility FAIR principle remains enforced when source data are no longer available? 

\section*{Acknowledgements}
E. Scholly's PhD 
is funded by BIAL-X\footnote{\url{https://www.bial-x.com/}}.
P.N. Sawadogo's PhD is funded by the Auvergne-Rhône-Alpes Region through the AURA-PMI project.
The HyperThesau project is funded by the
Laboratory of Excellence “Intelligence of Urban Worlds” (IMU)\footnote{\url{https://imu.universite-lyon.fr}}.


\bibliographystyle{ACM-Reference-Format}
\bibliography{references}

%
% ...

\end{document}